\documentclass[letterpaper]{article} % DO NOT CHANGE THIS
\usepackage{aaai25}  % DO NOT CHANGE THIS
\usepackage{times}  % DO NOT CHANGE THIS
\usepackage{helvet}  % DO NOT CHANGE THIS
\usepackage{courier}  % DO NOT CHANGE THIS
\usepackage[hyphens]{url}  % DO NOT CHANGE THIS
\usepackage{graphicx} % DO NOT CHANGE THIS
\usepackage{natbib}  % DO NOT CHANGE THIS AND DO NOT ADD ANY OPTIONS TO IT
\usepackage{caption} % DO NOT CHANGE THIS AND DO NOT ADD ANY OPTIONS TO IT
\usepackage{algorithm}

\usepackage{amsthm}
\usepackage{subfig}

\usepackage{algpseudocode}
\usepackage{amsmath}
\usepackage{amssymb}
\usepackage{booktabs}
\usepackage{newfloat}
\usepackage{listings}
\DeclareCaptionStyle{ruled}{labelfont=normalfont,labelsep=colon,strut=off} % DO NOT CHANGE THIS
\floatstyle{ruled}
\newfloat{listing}{tb}{lst}{}
\floatname{listing}{Listing}
\title{Learned Image Transmission with Hierarchical Variational Autoencoder}
\author{
    Guangyi Zhang\equalcontrib,
    Hanlei Li\equalcontrib,
    Yunlong Cai,
    Qiyu Hu,
    Guanding Yu,
    Runmin Zhang    
}
\affiliations{
    College of Information Science and Electronic Engineering, Zhejiang University, Hangzhou 310027, China \\
    \{zhangguangyi, hanleili, ylcai, qiyhu, yuguanding, runmin\_zhang\}@zju.edu.cn

}

\usepackage{bibentry}
\begin{document}
	\maketitle
	\begin{abstract}
		In this paper, we introduce an innovative hierarchical joint source-channel coding (HJSCC) framework for image transmission, utilizing a hierarchical variational autoencoder (VAE). Our approach leverages a combination of bottom-up and top-down paths at the transmitter to autoregressively generate multiple hierarchical representations of the original image. These representations are then directly mapped to channel symbols for transmission by the JSCC encoder. We extend this framework to scenarios with a feedback link, modeling transmission over a noisy channel as a probabilistic sampling process and deriving a novel generative formulation for JSCC with feedback. Compared with existing approaches, our proposed HJSCC provides enhanced adaptability  by dynamically adjusting transmission bandwidth, encoding these representations into varying amounts of channel symbols. Additionally, we introduce a rate attention module to guide the JSCC encoder in optimizing its encoding strategy based on prior information. Extensive experiments on images of varying resolutions demonstrate that our proposed model outperforms existing baselines in rate-distortion performance and maintains robustness against channel noise.
	\end{abstract}
	
	\section{Introduction}
	To meet the transmission requirements of heavy data traffic in future sixth-generation (6G) networks, wireless edge devices need to be equipped with higher transmission efficiency. Most contemporary systems employ a two-step strategy for data transmission: first, the raw data is compressed using a source codec, such as JPEG \cite{Wallace_TCE1992} and BPG \cite{Lainema_TCSVT2012}. Then, the encoded bits are protected with redundancy introduced by a carefully designed channel codec, such as LDPC and Polar codes \cite{Arikan_TIT2009}. However, in many practical applications, the bit length is generally finite, making it impossible to guarantee optimality. In this context, joint source-channel coding (JSCC) has emerged as a potential solution, offering higher coding gains than the traditional separation-based coding paradigm.
	
	With the revolutionary progress of deep learning in various fields, such as image compression \cite{balle_arxiv2018, Dailan_CVPR2021,Tongda_NIPS2023} and generative models \cite{Diederik_2014Arxiv, Razavi_NIPS2019}, a novel design paradigm for JSCC, called learned image transmission (LIT), has been conceived by formulating the communication pipeline as an end-to-end deep learning model \cite{Eirina_TCCN2019,Guangyi_TCOM2024, Lunan_JSAC2023}.  Specifically, these methods leverage powerful neural networks to implement the encoding and decoding processes. In this approach, the whole system is viewed as an autoencoder (AE), which can be jointly learned in a data-driven manner. A notable method proposed by \cite{Eirina_TCCN2019} employed CNNs to construct the source and channel codecs for wireless image transmission, achieving great performance by mapping the input image directly into channel symbols. Moreover, \citeauthor{deepjsccf, HaotianWu} investigate the JSCC with feedback. In this context, the transmission of these representations is divided into multiple phases, with the transmitter receiving the channel symbol vector after each phase, which simplifies the encoding process and improves the overall performance.

	Beyond deterministic AEs, some studies have employed variational autoencoders (VAE) to design JSCC systems \cite{Kristy_ICML2019, Yufei_Arxiv2023,vqvae}, where the channel symbols are generated through sampling. Part of these VAE-based methods show superior performance compared to deterministic AE-based methods, particularly under severe channel conditions.
	Though VAE-based methods have demonstrated remarkable performance, they experience significant performance degradation on high-resolution images. Furthermore, most existing methods only support fixed-rate coding, which contrasts with emerging works on transform coding-based image compression \cite{balle_arxiv2018}, where the compression rate for each image is determined by the estimated entropy of its feature representation and varies with different samples. Consequently, these methods are less flexible and adaptive, potentially leading to performance penalties. 
	%In response to this issue, \cite{electronics12224637} train a policy network with Gumbel-Softmax strategy that that reduces channel bandwidth utilization for high SNR scenarios or simple image contents. Moreover, the authors of \cite{Jincheng_2023JSAC} develops JSCC by controlling the transmission rate according to the estimated entropy. 
	
	In this work, we aim to overcome the limitations of previous methods while enhancing performance. Specifically, we develop a hierarchical JSCC (HJSCC) framework based on a powerful hierarchical VAE architecture \cite{Rewon_2020Arxiv}. Our transmitter employs both bottom-up and top-down paths to autoregressively generate multiple hierarchical representations of the original image. These representations are then mapped to channel symbols using multiple JSCC encoder blocks. 
	Building upon this, we further explore the application of HJSCC in a classical scenario where a feedback link exists. By modeling transmission over a noisy channel as a probabilistic sampling process, we derive a novel generative formulation for JSCC with feedback, which achieves significantly better performance than most existing advanced schemes.
	While there have been attempts at variable-rate transmission \cite{Jincheng_2023JSAC, electronics12224637,PredictiveJSCC,Mingyu} in the realm of JSCC without feedback, the problem of rate-adaptive design for JSCC with feedback remains underexplored. Unlike existing works \cite{deepjsccf, HaotianWu,harq1,harq2}, we leverage the prior distribution (which characterizes the entropy information) of each representation to generate masks that control the number of symbols for each representation. This approach allows us to dynamically adjust the transmission rate.
	Additionally, we introduce a rate attention module to guide the JSCC encoder in adjusting the encoding strategy according to its prior information. 
	
	In summary, our contributions are as follows:
	\begin{itemize}
		\item \textbf{HJSCC Framework}: Developing a hierarchical   scheme that is able to support the transmission of high-resolution images.
		\item \textbf{HJSCC with Feedback}: Extending HJSCC to the case with feedback, by viewing the transmission as a sampling process and deriving a generative formulation.
		\item \textbf{Dynamic Rate Control:} By utilizing the entropy information of representations to dynamically control the transmission rate, this approach bridges the gap of lacking rate-adaptive design when a feedback link is present.
		\item \textbf{Rate Attention Module:} Proposing a spatial grouping strategy and a rate attention module to improve the overall rate-distortion performance.
		\item \textbf{Experimental Studies:} Providing substantial experiments to verify the effectiveness of the proposed method, demonstrating that the proposed scheme achieves better coding gain than emerging deep learning-based JSCC and separation-based digital transmission schemes. 
	\end{itemize}

	\section{Related Works}
	\label{gen_inst}

	\paragraph{Varational Autoencoder}
	VAE can be employed as deep generative models capable of generating high-dimensional data based on a low-dimensional latent space, and is a variant of autoencoder \cite{Casper_NIPS2016, NEURIPS2020_Cemgil}. By sampling from the learned latent space distribution and passing these samples through the decoder network, VAE can generate new data points that resemble the training data. However, the original VAE is known to perform worse than many other generative models, particularly when applied to high-resolution images. \cite{Vahdat_NIPS,Rewon_2020Arxiv,Yuhta_2024Arxiv} addressed this by proposing a deep hierarchical VAE, where the latent variable is divided into several disjoint groups, achieving significantly better performance than standard VAEs. More recently, VAE has been applied to compression tasks \cite{Duan_WACV2023,Ming_AAAI2024, James_ICLR,Kingma_PMLR}.

	\paragraph{Learned Image Transmission}
	Unlike the separation-based design described above, recent studies have delved into the utilization of AE and its variants, e.g., VAE to design wireless image transmission systems, resulting in a number of efficient methods \cite{Eirina_TCCN2019,Jialong_2022TCSVT,Guangyi_TCOM2024, Lunan_JSAC2023, Jialong_2022TCSVT,Yang_Arxiv2023, Saidutta_ISIT2019}. In particular, \cite{Eirina_TCCN2019,Haotian_TWC} and \cite{Yang_Arxiv2023} conceived of using neural networks to simultaneously finish the source encoding/decoding and channel coding/decoding,  with the goal of jointly optimizing the entire system to maximize PSNR. 
	The VAE-based methods adopted probabilistic modeling, wherein the encoding process is characterized as a stochastic procedure. In these systems, channel symbols are generated by sampling from the probability distribution conditioned on the input image \cite{Yufei_Arxiv2023,Saidutta_ISIT2019}. These approaches have demonstrated superior performance, especially under severe transmission conditions.

	\section{Proposed Methods}
	
	\subsection{Background}

	\begin{figure}
		\begin{centering}	
			\includegraphics[width=0.46 \textwidth]{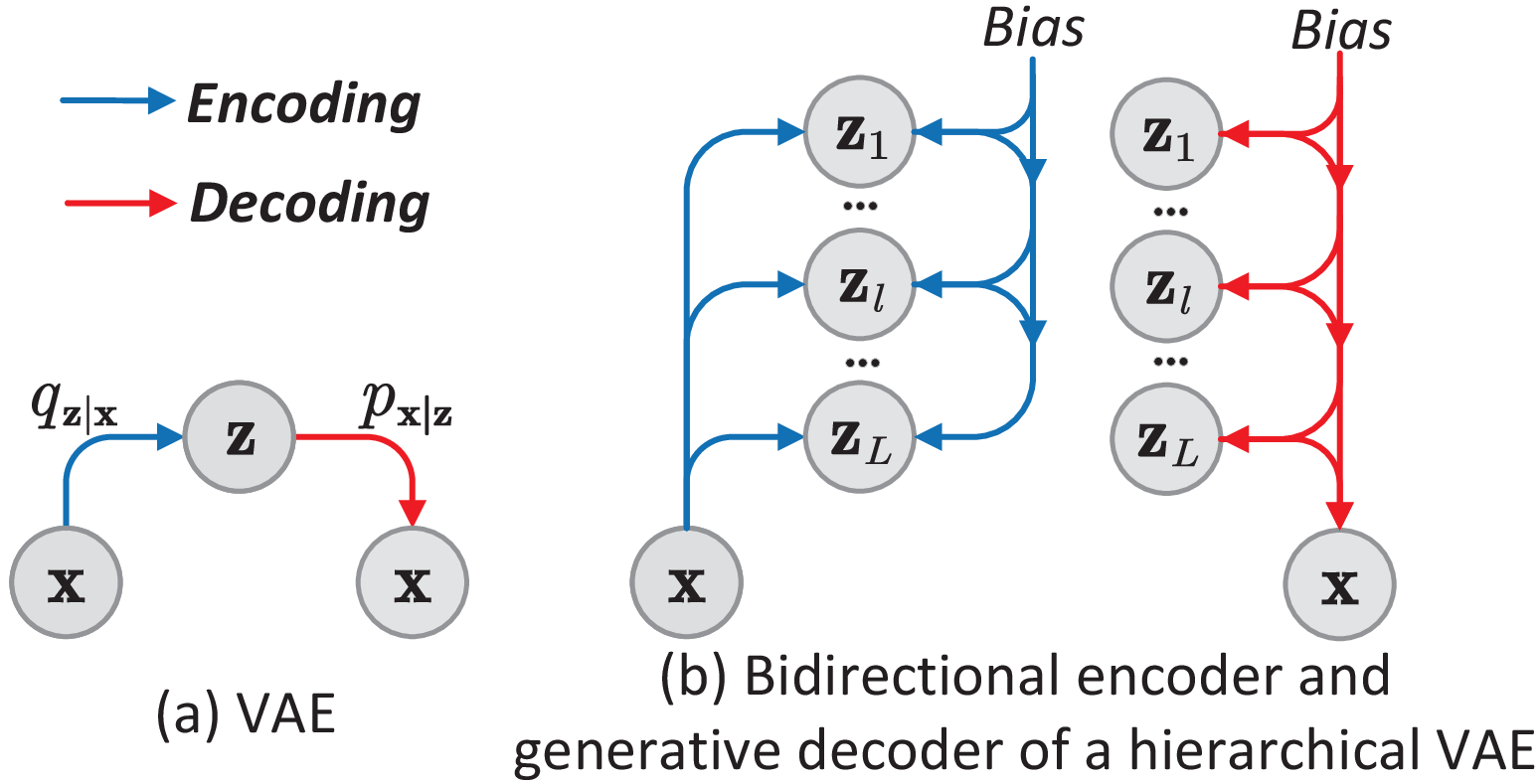}
			\par \end{centering}
		\caption{Probabilistic model of VAEs and hierarchical ResNet VAE. The bias is a trainable parameter.}
		\label{VAEHVAE}
	\end{figure}

	\subsubsection{VAE and Hierarchical VAE} \label{HVAE_Section}
	As stated in \cite{NEURIPS2020_Cemgil}, VAE is a stochastic variational inference scheme that can be applied to various intelligent tasks, such as recognition, denoising, and generation. As shown in Fig. \ref{VAEHVAE}(a), to formulate a vision-related model, we typically start with the following premises.  Let $\mathbf{x}$ denote an image intensity vector, which is drawn from a dataset $\mathcal{X}$ with distribution $p_{\mathbf{x}}$. Another variable is the latent variable $\mathbf{z}$, with a prior $p_\mathbf{z}$. The main target of a VAE is to learn a generative model (decoder) $p_{\mathbf{x}|\mathbf{z}}$ for sampling, and a posterior density model (encoder) $q_{\mathbf{z}|\mathbf{x}}$ for variational inference. The objective for learning a VAE model can be formulated as minimizing the (variational) upper bound on the marginal likelihood of a batch of data points, as given by
	\begin{equation}\label{VAE_loss}
		\mathcal{L}\left(\boldsymbol{\theta}, \boldsymbol{\phi} \right)=\mathbb{E}_{\mathbf{x}\sim p_{\mathbf{x}},\mathbf{z}\sim q_{\mathbf{z}|\mathbf{x}} }\big[D_{K L}\left(q_{\mathbf{z} | \mathbf{x}} \| p_{\mathbf{z}}\right)-\log p_{\mathbf{x} | \mathbf{z}}\big],
	\end{equation}
	where $D_{K L}\left(q_{\mathbf{z} | \mathbf{x}} \| p_{\mathbf{z}}\right)=\log\big(q_{\mathbf{z} | \mathbf{x}} / p_{\mathbf{z}}\big)$, $\boldsymbol{\theta}$ and $\boldsymbol{\phi}$ represent the parameters of the encoder  $q_{\mathbf{z}|\mathbf{x}}$ and decoder $p_{\mathbf{x}|\mathbf{z}}$, respectively. 
	
	Hierarchical VAEs are a series of VAE models that partition the latent variables into several disjoint groups. The probabilistic diagram of a classical hierarchical VAE, the ResNet VAE, is shown in Fig. \ref{VAEHVAE}(b), consisting of a bidirectional encoder and a generative decoder. Specifically, the latent variables can be denoted by $\mathbf{z} \triangleq\left\{\mathbf{z}_1, \mathbf{z}_2, \ldots, \mathbf{z}_L\right\}$, where $L$ represents the number of groups. The prior for $\mathbf{z}$ is modeled as $p_{\mathbf{z}_{1: L}} = \prod_l p_{\mathbf{z}_l | \mathbf{z}_{<l}}$, and the approximate posterior is denoted as $q_{\mathbf{z}} = \prod_l q_{\mathbf{z}_l | \mathbf{z}_{<l}, \mathbf{x}}$, where $\mathbf{z}_{<l}$ represents $\left\{\mathbf{z}_1, \mathbf{z}_2, \ldots, \mathbf{z}_{l-1}\right\}$. In general, the dimension of $\mathbf{z}_l$ is designed to be smaller than that of $\mathbf{z}_{l+1}$, fulfilling the target of capturing the coarse-to-fine nature of images. The objective for training a hierarchical VAE mode can be obtained by extending Eq. (\ref{VAE_loss}) for multiple latent variables, as given by
	\begin{equation}\label{HVAE_loss}
		\begin{aligned}
			& \mathcal{L}\left(\boldsymbol{\theta}, \boldsymbol{\phi} \right)=\\
			&\quad\mathbb{E}_{\mathbf{x}\sim p_{\mathbf{x}},\mathbf{z}\sim q_{\mathbf{z} |\mathbf{x}} }  \left[  \sum_{l=1}^{L} D_{K L}\left(q_{\mathbf{z}_l | \mathbf{z}_{<l} ,\mathbf{x}} \| p_{\mathbf{z}_l | \mathbf{z}_{<l}}\right)\!-\!\log p_{\mathbf{x} | \mathbf{z}} \right],
		\end{aligned}
	\end{equation}
	where we define $\mathbf{z}_{<1}$ as an empty set, and thus $p_{\mathbf{z}_1 | \mathbf{z}_{<1}} = p_{\mathbf{z}_1}$ and $q_{\mathbf{z}_1| \mathbf{z}_{<1} ,\mathbf{x}} = q_{\mathbf{z}_1| \mathbf{x}}$.

	\begin{figure}
		\begin{centering}	
			\includegraphics[width=0.45 \textwidth]{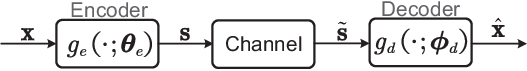}
			\par \end{centering}
		\caption{Diagram of a deep learning-based JSCC system.}
		\label{FrameworkJSCC}
	\end{figure}
	
	\subsection{Proposed HJSCC}
	\subsubsection{System Overview}\label{overview}
	Here, we aim to give a brief overview of deep learning-based JSCC. In particular, the model of a typical JSCC is shown in Fig. \ref{FrameworkJSCC}. The transmitter employs a JSCC encoder to map the input image $\mathbf{x} \in \mathbb{R}^{N}$ directly into channel symbol vector $\mathbf{s} \in \mathbb{C}^{K}$ for transmission, where $N$ denotes the number of pixels and $K$ represents the number of channel symbols. This process can be expressed as $\mathbf{s}=g_e(\mathbf{x}; \boldsymbol{\theta}_e)$, where $g_e$ signifies the encoding function and $\boldsymbol{\theta}_e$ represents its trainable parameters. Accounting for the limited transmission power, the transmitted signal should satisfy the power constraint $P$, implying that  $\|\mathbf{s}\|^2_2/K \leq P$. Subsequently, the channel symbol vector is transmitted through the wireless channel, as given by $\mathbf{\tilde{s}}=\mathbf{s}+\mathbf{n}$, where $\mathbf{n}\sim \mathcal{N}(0, \sigma^2_n\mathbf{I})$ is additive white Gaussian noise (AWGN). At the receiver, the received noisy signal $\mathbf{\tilde{s}}$ is processed by a decoder $g_d$ to obtain the reconstructed image $\mathbf{\bar{x}}=g_d(\mathbf{\tilde{s}};\boldsymbol{\phi}_d)$, with  $\boldsymbol{\phi}_d$ denoting the trainable parameters of the JSCC decoder. The optimization objective of a deep learning-based JSCC system is to minimize the difference between $\mathbf{x}$ and $\mathbf{\bar{x}}$, and thus mean-square error (MSE) can be employed as the loss function. Furthermore, to evaluate the performance of a JSCC system and ensure fairness, we define the  \textbf{signal-to-noise ratio (SNR)} as $\text {SNR }=10 \log \frac{P}{\sigma^2}(\mathrm{dB})$, which characterizes the channel quality of the system. Then, we introduce the \textbf{channel bandwidth ratio (CBR)} to describe the \textbf{transmission rate} (overhead), which is expressed as $\text{CBR}=K/N$. Intuitively, CBR actually signifies the number of symbols for transmitting one pixel, and \textit{a higher CBR brings a higher overhead, while usually  resulting in a better system performance.}
	
\begin{figure*}[t]
		\begin{centering}
			\includegraphics[width=0.8 \textwidth]{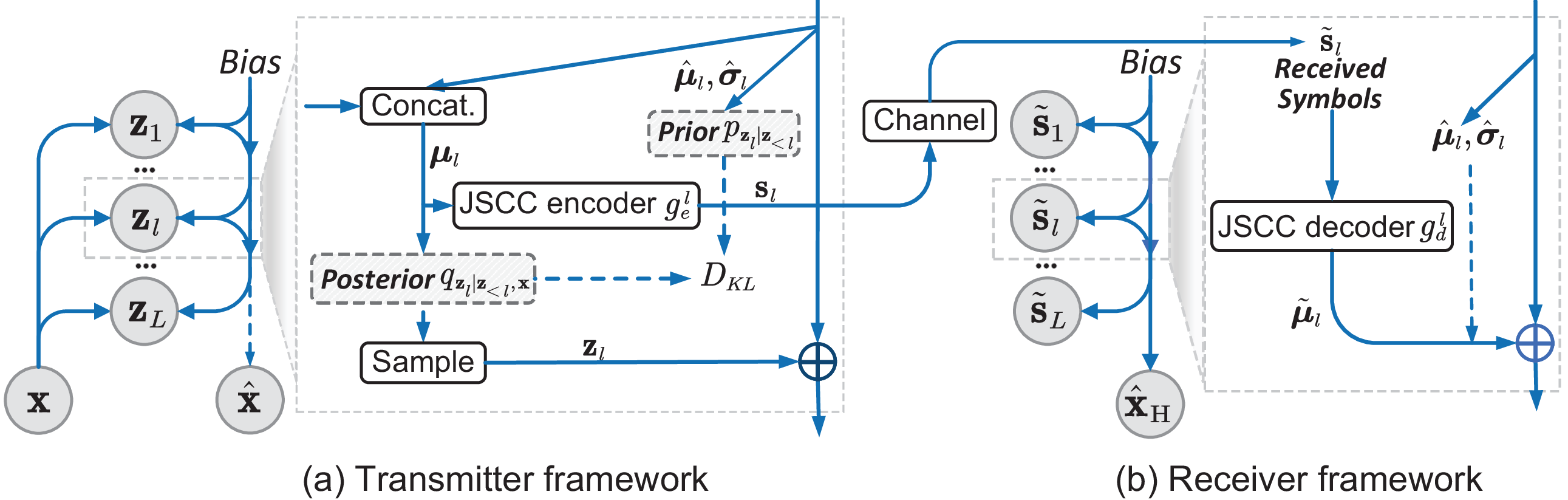}
			\par \end{centering}
		\caption{The probabilistic diagram of the proposed HJSCC. The transmitter employs the bottom-up and top-down paths for encoding, while the receiver reconstructs the image with the received symbols.}
		\label{HJSCCFramework}
	\end{figure*}

	While previous studies have achieved excellent rate-distortion performance based on the framework depicted in Fig. \ref{FrameworkJSCC}, it is evident that the transmission rate is solely determined by the image resolution. This limitation can result in performance degradation in overall rate-distortion due to the inability to adaptively adjust the rate for each image \cite{He_2022_CVPR}.
	To address this issue, we propose our HJSCC framework, as illustrated in Fig. \ref{HJSCCFramework}.
	Given an image $\mathbf{x}$, the bottom-up path generates a set of latent features, which are subsequently passed to the top-down path to autoregressively generate the 
	latent representations $\boldsymbol{\mu} \triangleq\left\{\boldsymbol{\mu}_1, \boldsymbol{\mu}_2, \ldots, \boldsymbol{\mu}_L\right\}$. These representations are then fed to a set of JSCC encoders $g_e=\left\{ g^1_e,g^2_e,\ldots,g_e^L  \right\}$, respectively. In this way, we are able to obtain a set of channel symbol vectors $\mathbf{s} \triangleq\left\{\mathbf{s}_1, \mathbf{s}_2, \ldots, \mathbf{s}_L\right\}$, where $\mathbf{s}_l=g_e^l(\mathbf{z}_l)$. Then, these channel symbol vectors are transmitted to the receiver through the wireless link, and the received symbol vectors are represented by $\mathbf{\tilde{s}} \triangleq\left\{\mathbf{\tilde{s}}_1, \mathbf{\tilde{s}}_2, \ldots, \mathbf{\tilde{s}}_L\right\}$. At the receiver, the noisy $\mathbf{\tilde{s}}$ undergoes processing by the JSCC decoder $g_d^l$, and we obtain $\boldsymbol{\tilde{\mu}} \triangleq\left\{\boldsymbol{\tilde{\mu}}_1, \boldsymbol{\tilde{\mu}}_2, \ldots, \boldsymbol{\tilde{\mu}}_L\right\}$. With $\boldsymbol{\tilde{\mu}}$ at hand, the receiver can reconstruct the image using the top-down path (decoder), and the reconstructed image is denoted by $\mathbf{\hat{x}}_\text{H}$. Our objective is to minimize the distortion between the transmitted image $\mathbf{x}$ and $\mathbf{\hat{x}}_\text{H}$. 
	
	\begin{figure*}[t]
		\begin{centering}
			\includegraphics[width=1.0 \textwidth]{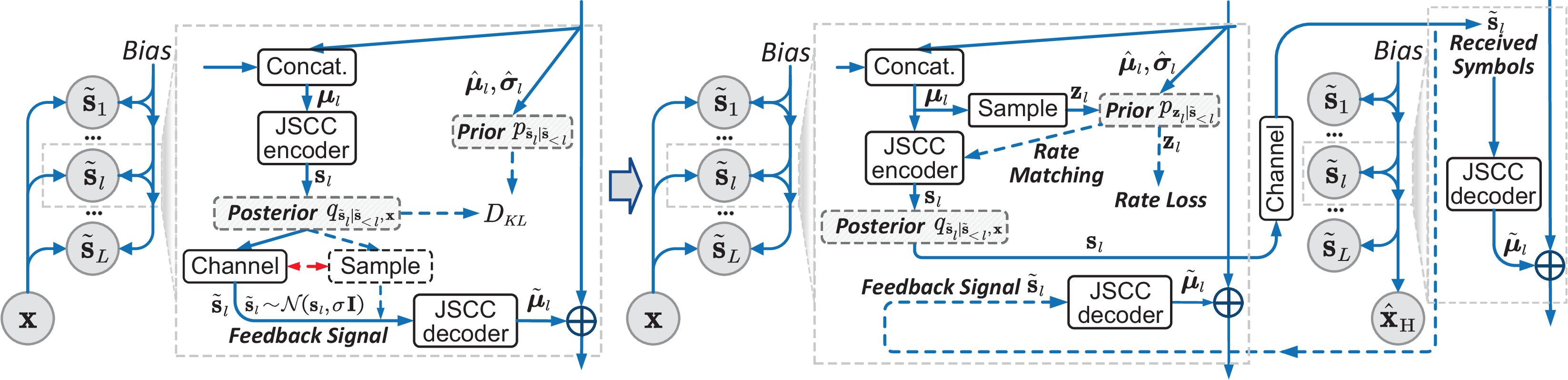}
			\par \end{centering}
		\caption{The probabilistic diagram of the proposed HJSCC with feedback. }
		\label{HJSCCFrameworkFeedback}
	\end{figure*}
	
	\subsection{Training Objective Formulation}
	We aim to develop a rate-adaptive JSCC model capable of adjusting the transmission rate, CBR, based on the source content. To this end, we propose an inherited training strategy motivated by \cite{Jincheng_2023JSAC}, where the objective of minimizing $d(\mathbf{x}, \mathbf{\hat{x}}_\text{H})$ is guided by a learned image coder. Specifically, we have designs for the posteriors, priors, and training objectives as follows \cite{Duan_WACV2023}.
	
	\textbf{Posteriors}: The posteriors $q_{\mathbf{z}_l | \mathbf{z}_{<l} ,\mathbf{x}}$  is set to a uniform distribution \begin{equation}
		q_{\mathbf{z}_l | \mathbf{z}_{<l} ,\mathbf{x}}(\mathbf{z}_l | \mathbf{z}_{<l} ,\mathbf{x}) \triangleq \prod_i\mathcal{U}\left(\mu_l^{(i)}-\tfrac{1}{2}, \mu_l^{(i)}+\tfrac{1}{2}\right),
	\end{equation}
	where $\mathcal{U}$ denotes a uniform distribution centered on $\mu_l^{(i)}$, and $\mu_l^{(i)}$ is the $i$-th element of parameter  $\boldsymbol{\mu}_l$, which is obtained from the $l$-th posterior branch.
	
	\textbf{Priors}: The conditional prior distribution $p_{\mathbf{z}_l | \mathbf{z}_{<l}} $ is  defined to be a Gaussian distribution convolved with a uniform distribution \cite{balle_arxiv2018}:
	\begin{equation}
		p_{\mathbf{z}_l | \mathbf{z}_{<l}} (\mathbf{z}_l | \mathbf{z}_{<l}) \triangleq \prod_i\mathcal{N}\left(\hat{\mu}_l^{(i)}, (\hat{\sigma}_l^{(i)})^2\right) * \mathcal{U}\left(-\tfrac{1}{2}, \tfrac{1}{2}\right) ,
	\end{equation}
	where $*$ represents the convolution operation.
	
	\textbf{Training Loss:}
	With the posteriors and priors defined above, the loss function (\ref{HVAE_loss}) for an image compression model \cite{Duan_WACV2023} can be expressed as 
	\begin{equation}\label{QARVLoss}
		\begin{aligned}
			\mathcal{L} 
			%& =\mathbb{E}_{\mathbf{x}\sim p_{\mathbf{x}},\mathbf{z}\sim q_{\mathbf{z} |\mathbf{x}} }  \left[ \sum_{l=1}^{L} D_{K L}\left(q_{\mathbf{z}_l | \mathbf{z}_{<l} ,\mathbf{x}} \| p_{\mathbf{z}_l | \mathbf{z}_{<l}}\right)-\log p_{\mathbf{x} \mid \mathbf{z}}(\mathbf{x} | \mathbf{z})\right]\\
			& =\mathbb{E}_{\mathbf{x}\sim p_{\mathbf{x}},\mathbf{z}\sim q_{\mathbf{z} |\mathbf{x}} }\left[\sum_{l=1}^L \log \frac{1}{p_{\mathbf{z}_l | \mathbf{z}_{<l}}(\mathbf{z}_l | \mathbf{z}_{<l})}+\lambda \cdot d(\mathbf{x}, \mathbf{\hat{x}})\right],
		\end{aligned}
	\end{equation}
	where $\lambda$ is the introduced weight to control the tradeoff between rate (the first term) and distortion (the second term) $d(\mathbf{x}, \mathbf{\hat{x}})$. Thus, (\ref{QARVLoss}) can be utilized for image compression to achieve a trade-off between rate and distortion.
	Under the guidance of (\ref{QARVLoss}), we further develop the optimization problem for HJSCC as follows:
	\begin{equation}
		\begin{aligned}
			\mathcal{L}	
			&=  \mathbb{E}_{\mathbf{x}\sim p_{\mathbf{x}},\mathbf{z}\sim q_{\mathbf{z} |\mathbf{x}} }\big[ \\
			&\quad \underbrace{\sum_{l=1}^L - \alpha \log p_{\mathbf{z}_l | \mathbf{z}_{<l}}(\mathbf{z}_l | \mathbf{z}_{<l})}_{\text{transmission rate}}+\lambda \cdot (\underbrace{d(\mathbf{x}, \mathbf{\hat{x}}) +  d(\mathbf{x}, \mathbf{\hat{x}}_{\text{H}}))}_{\text{weighted distortion}}\big], 
			%&=  \mathbb{E}_{\mathbf{x}\sim p_{\mathbf{x}},\mathbf{z}\sim q_{\mathbf{z} |\mathbf{x}} }\big[\sum_{l=1}^L \sum_{j=1}- \alpha \log p_{z_l^{(j)} | \mathbf{z}_{<l}}(z_l^{(j)} | \mathbf{z}_{<l})+\\
			%&\quad \;\lambda \cdot (d(\mathbf{x}, \mathbf{\hat{x}}) +  d(\mathbf{x}, \mathbf{\hat{x}}_{\text{H}}))\big], \\
		\end{aligned}
	\end{equation}
	where the calculation ways of each variable can be found in Fig. \ref{HJSCCFramework}.
	Intuitively, there are two main differences from the loss function (\ref{QARVLoss}). The first is the additional distortion term to optimize the transmission distortion $d(\mathbf{x},\mathbf{\hat{x}_{\text{H}}})$, which helps to obtain a robust representation against noise. The second is the scaling parameter $\alpha$ to control the relation between the entropy of the latent $\mathbf{z}_l$ and the transmission rate for $\mathbf{s}_l$. 
	
	To flexibly adjust the CBR for different images, a natural choice for reference is the prior of latent variable $\mathbf{z}_l$ that is correlated to the bit length after entropy coding, where we assume the entropy of $\mathbf{z}_l$ is positively related to the entropy of $\boldsymbol{\mu}_l$.  Thus, the CBR for encoding $\boldsymbol{\mu}_l$ is designed to be proportional to the prior $p_{\mathbf{z}_l | \mathbf{z}_{<l}}$. Moreover, noting that the channel bandwidth is determined by reducing the number of the channel symbols, we achieve rate adjustment by masking the channel symbol vector according to the transmission rate $\alpha p_{\mathbf{z}_l | \mathbf{z}_{<l}}$, which will be introduced in the following sections.

	\begin{figure*}[t]
		\begin{centering}
			\includegraphics[width=0.93 \textwidth]{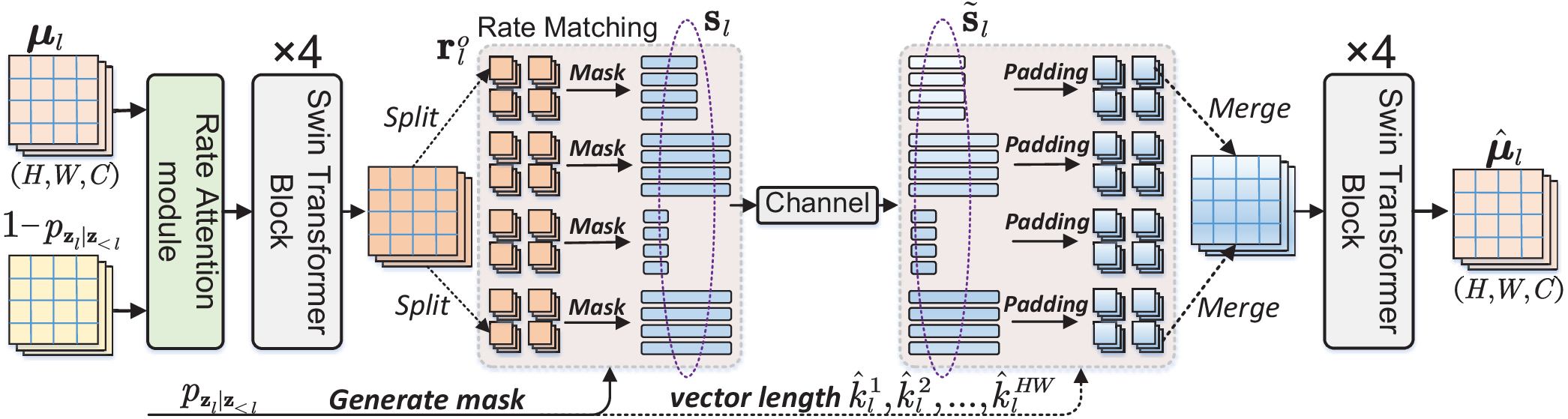}
			\par \end{centering}
		\caption{The illustration of the process of the JSCC encoder and JSCC decoder for transmitting latent representation $\boldsymbol{\mu}_l$.}
		\label{JSCCencoder}
	\end{figure*}
	
	\subsection{HJSCC with Feedback}
	Building upon previous designs, we further investigate the JSCC scenario with feedback link from the receiver to the transmitter. In this scenario, image transmission is divided into multiple phases, allowing the transmitter to use the received channel symbol vectors from previous phases when encoding channel symbols in the current phase.
	
	As shown in the left framework of Fig. \ref{HJSCCFrameworkFeedback}, we formulate our HJSCC with feedback by viewing the received symbols vector $\mathbf{\tilde{s}}_l$ as the latent variable. In this case, the loss function for training HJSCC with feedback can be written as
	\begin{equation}\label{QARVLossFeedback}
		\begin{aligned}
			\mathcal{L} 
			& =\mathbb{E}_{\mathbf{x}\sim p_{\mathbf{x}},\mathbf{\tilde{s}}\sim q_{\mathbf{\tilde{s}} |\mathbf{x}} }  \left[ \sum_{l=1}^{L} \log \frac{q_{\mathbf{\tilde{s}}_l | \mathbf{\tilde{s}}_{<l},\mathbf{x}}(\mathbf{\tilde{s}}_l | \mathbf{\tilde{s}}_{<l},\!\mathbf{x})}{p_{\mathbf{\tilde{s}}_l | \mathbf{\tilde{s}}_{<l}}(\mathbf{\tilde{s}}_l | \mathbf{\tilde{s}}_{<l})}\!-\!\log p_{\mathbf{x} \mid \mathbf{\tilde{s}}}\right].
		\end{aligned}
	\end{equation}
	Intuitively, we view the transmission over noisy channels as a process of sampling. In particular, since we consider AWGN channels, i.e., $\mathbf{\tilde{s}}_l=\mathbf{s}_l + \mathbf{n}_l$ ,the posteriors $q_{\mathbf{s}_l | \mathbf{s}_{<l} ,\mathbf{x}}$  is actually a Gaussian distribution \begin{equation}
		q_{\mathbf{\tilde{s}}_l | \mathbf{\tilde{s}}_{<l} ,\mathbf{x}}(\mathbf{\tilde{s}}_l | \mathbf{\tilde{s}}_{<l} ,\mathbf{x}) \triangleq \prod_i\mathcal{N}\left(s_l^{(i)}, \sigma_n^{2}\right),
	\end{equation}
	where $\mathbf{s}_l$ can be directly computed with known $\mathbf{x}$ and $\mathbf{\tilde{s}}_{<l}$.
	Besides, with this formulation, the generation of the $l$-th channel symbol vector $\mathbf{\tilde{s}}_l$ is conditioned on the  received vectors in the former phases. It enables the HJSCC to adjust the transmitted symbols based on feedback signals. 
	In this way, we also fortunately find that the term $\log q_{\mathbf{\tilde{s}}_l | \mathbf{\tilde{s}}_{<l},\mathbf{x}}(\mathbf{\tilde{s}}_l | \mathbf{\tilde{s}}_{<l},\!\mathbf{x})$ in (\ref{QARVLossFeedback}) is only related to the channel noise and will not introduce gradient to the whole model. Thus, this term can be directly dropped from the loss function. 
	
	We aim to achieve rate-adaptive transmission for HJSCC in the presence of feedback link. Similar to the scenario without feedback, we are able to take the prior information $p_{\mathbf{\tilde{s}}_l | \mathbf{\tilde{s}}_{<l}}(\mathbf{\tilde{s}}_l | \mathbf{\tilde{s}}_{<l})$ as a rate indicator. However, the calculation of this prior depends on the values of $\mathbf{\tilde{s}}_l$, which is unknown before transmission. As shown in the right of Fig. \ref{HJSCCFrameworkFeedback}, we address this by proposing a new  prior $p_{\mathbf{z}_l | \mathbf{\tilde{s}}_{<l}}(\mathbf{z}_l | \mathbf{\tilde{s}}_{<l})$ as a substitution, where $\mathbf{z}_l$ is sampled from a uniform distribution centered on $\boldsymbol{\mu}_l$. This is inspired by the fact that $p_{\mathbf{z}_l | \mathbf{\tilde{s}}_{<l}}(\mathbf{z}_l | \mathbf{\tilde{s}}_{<l})$ is positively related to $p_{\mathbf{\tilde{s}}_l | \mathbf{\tilde{s}}_{<l}}(\mathbf{\tilde{s}}_l | \mathbf{\tilde{s}}_{<l})$, and thus can be employed as the rate term when training the model. As a result, the loss function for training the HJSCC with feedback can be written as 
	\begin{equation}\label{feedback}
		\begin{aligned}
			\mathcal{L} 
			& =\mathbb{E}_{\mathbf{x}\sim p_{\mathbf{x}},\mathbf{\tilde{s}}\sim q_{\mathbf{\tilde{s}} |\mathbf{x}} }  \left[ \sum_{l=1}^{L} \log \frac{\text{const}}{ p_{\mathbf{z}_l | \mathbf{\tilde{s}}_{<l}}\cdot \beta}\!+\lambda d(\mathbf{x}, \mathbf{\hat{x}}_{\text{H}})) \right],
		\end{aligned}
	\end{equation}
	where $\beta$ denotes the scaling factor.

	\subsection{Masking and Length Information Reduction}
	A visualized example of our proposed masking strategy is shown in Fig. \ref{JSCCencoder}. Particularly, we employ the Swin Transformer blocks to implement our JSCC encoder, since we find that this architecture presents better robustness against channel noise. In this way, the shape of the output feature at the $l$-th layer, $\mathbf{r}_l$, can be denoted as $C\times H \times W$, which is the same as that of $\boldsymbol{\mu}_l$. We split the output of the Swin Transformer blocks, $\mathbf{r}_l$, into $HW$ sequences $\mathbf{r}_l^o$, for $o=1,2,\ldots,HW$, where the length of each $\mathbf{r}_l^o$ is $C$. Then, we include a rate-matching layer after the Swin Transformer blocks. It accepts two inputs, $\mathbf{r}_l$ and $\alpha p_{\mathbf{z}_l | \mathbf{z}_{<l}}$, and generates the masked channel symbol vector $\mathbf{s}_l$. Specifically, the length for each vector $\mathbf{s}_l^o$ will be constrained to $k_l^o=\sum_{c=1}^{C}- \alpha \log p_{z_l^{(o,c)} | \mathbf{z}_{<l}}(z_l^{(o,c)} | \mathbf{z}_{<l})$, which actually represents the summation of entropy along all the $C$ dimensions of $\mathbf{z}^{o}_l$. We adjust the length by generating a mask vector $\mathbf{m}_l^o$, which can be written as 
	\begin{equation}
		\mathbf{m}_l^o=[\underbrace{1,1,\ldots,1}_{k_l^o},0,\ldots,0],
	\end{equation}
	indicating that only the former $k_l^o$ elements of $\mathbf{r}_l^o$ will be used as the channel symbols, i.e., $\mathbf{s}_l^o=\mathbf{r}_l^o \odot \mathbf{m}_l^o$, where $\odot$ denotes the element-wise multiplication. 
	
%	\begin{figure}[t]
%		\begin{centering}
%			\includegraphics[width=0.34 \textwidth]{Diagram/RateAttention.eps}
%			\par \end{centering}
%		\caption{Detailed illustrations of the rate attention module.}
%		\label{rateattention}
%	\end{figure}
	Though this enables the transmitter to determine the transmission rate adaptively for each image, this design introduces a length-matching issue. Specifically, the receiver needs to know the length of each transmitted symbol vector to identify the channel symbols from different spatial positions and layers. This requirement, however, adds an overhead of communicating this information to the receiver. To mitigate this overhead, we propose two practical designs:
	
	Firstly, instead of considering infinite precision, we opt for a finite set of length options, comprising $\{2^{N_q} \}$ integers, where $N_q$ is a selected integer. We define the optional length set as 
	$\mathcal{Q}=\{q_1,q_2,\ldots,q_{2^{N_q}}\}$. Then, for each $\mathbf{s}_l^o$, the transmission rate is quantilized to $\hat{k}_l^o=Q(k_l^o)=Q(\sum_{c=1}^{C}- \alpha \log p_{z_l^{(o,c)} | \mathbf{z}_{<l}}(z_l^{(o,c)} | \mathbf{z}_{<l}))$ with the optional length set $\mathcal{Q}$. Therefore, we incorporate an extra link to transmit $N_q$ bits as side information to inform the length for each $\mathbf{s}_l^o$, where we assume this information should be transmitted without errors. The additional overhead for each $\mathbf{s}_l^o$ is $\frac{\log_2N_q}{C_r}$, where $C_r$ denotes the channel capacity. 
	
	\begin{figure*}[t]
		\begin{centering}
			\subfloat[]{\label{f1}\includegraphics[width=5.93cm]{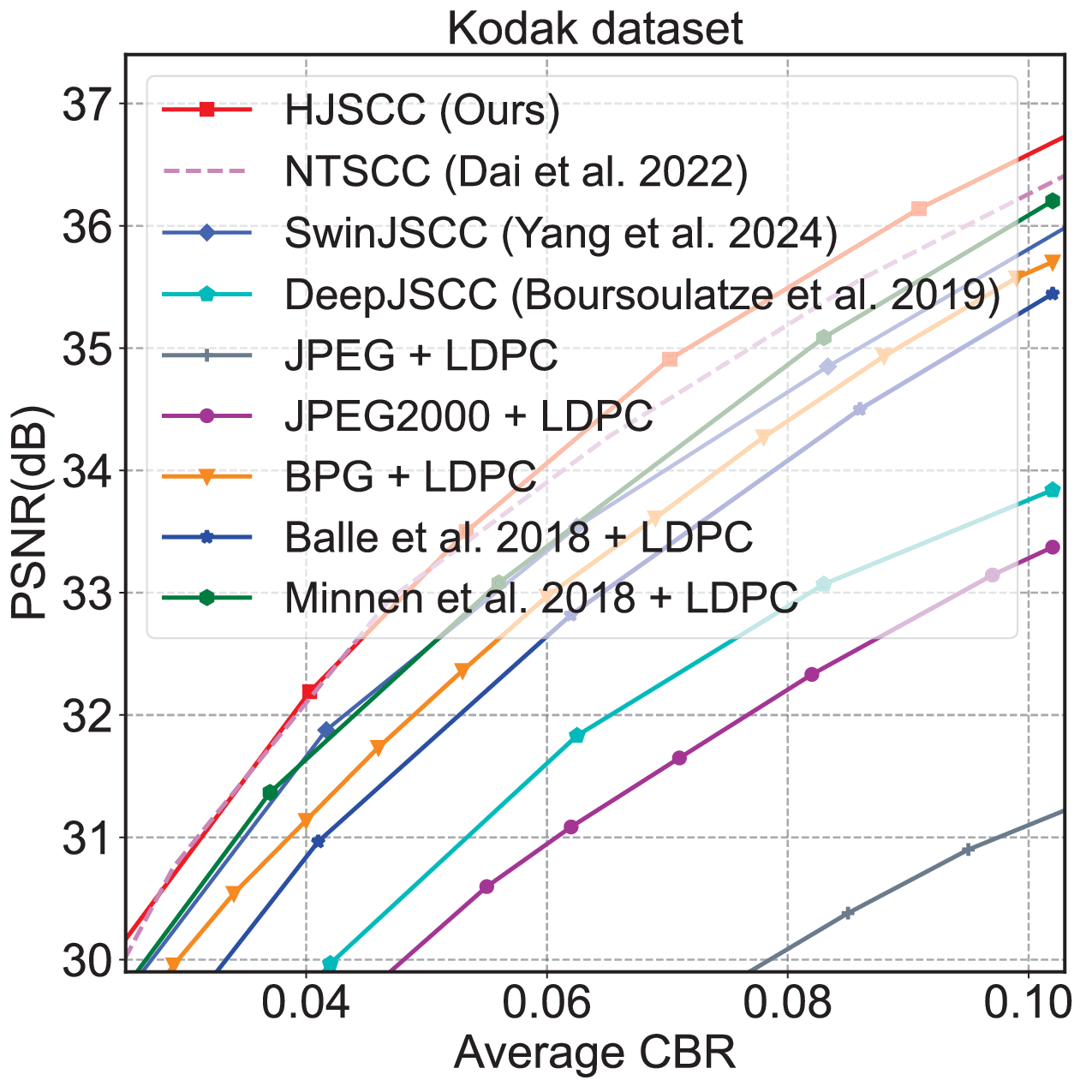}} 
			\subfloat[]{\label{f3}\includegraphics[width=5.93cm]{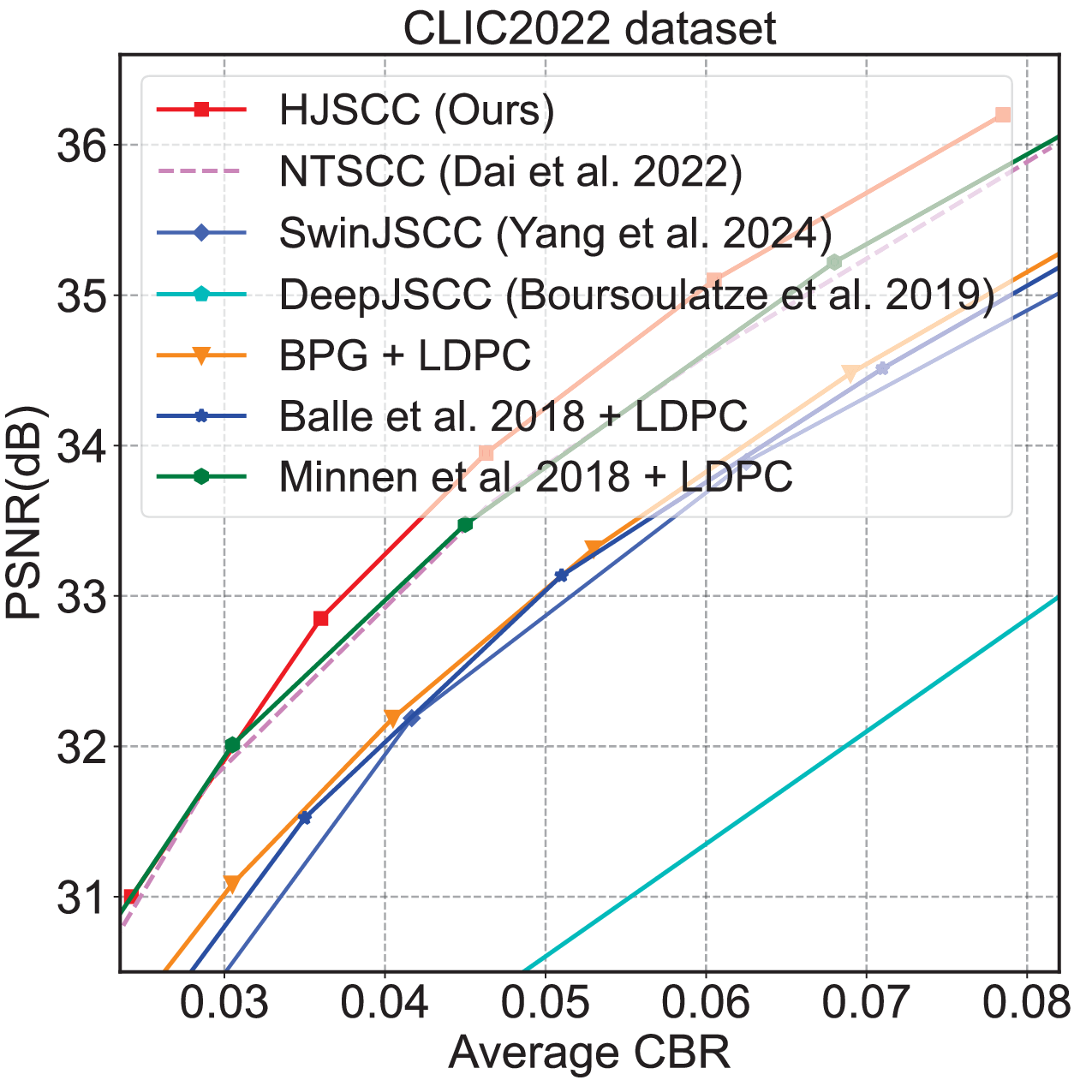}}
			\subfloat[]{\label{f4}\includegraphics[width=5.93cm]{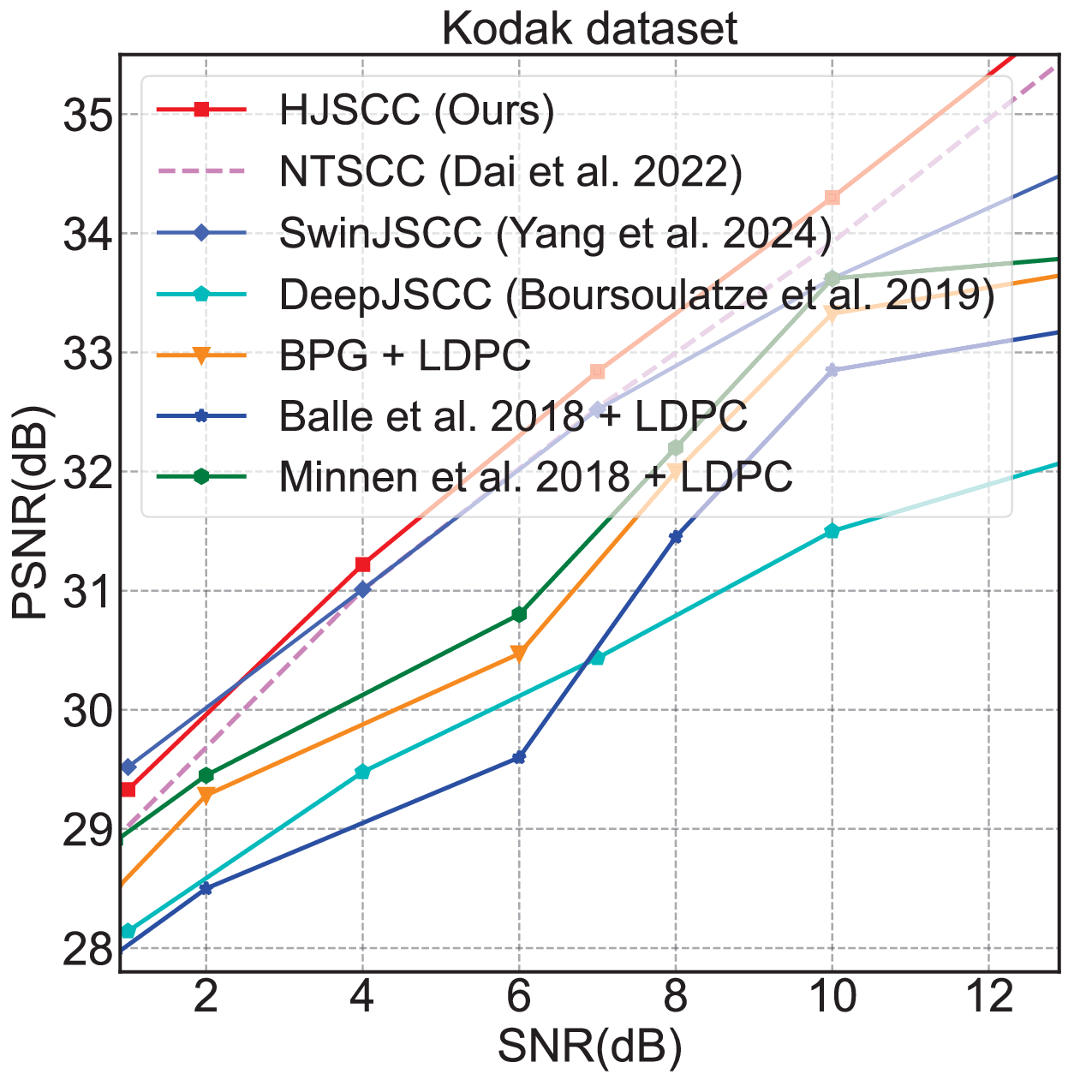}}
			\caption{The end-to-end distortion performance versus over different datasets. The results are evaluated on  (a) Kodak and (b) CLIC2022 datasets, at SNR $=10$ dB. (c) The end-to-end distortion performance on Kodak dataset versus the SNR.} 
			\label{PSNR_CBR}
		\end{centering}
	\end{figure*}
	Secondly, for each $\mathbf{s}_l$ we need in total $HW N_q$ extra bits as the side information. In comparison to the transmission overhead at the $l$-th layer, $\sum_{o=1}^{HW} \hat{k}_l^o$, this amount of side information will be quite significant when $C$ is small. To address this issue, we design a spatial grouping strategy. As shown in the middle part of Fig. \ref{JSCCencoder}, we split the $\mathbf{r}_l$ into multiple patches along the width and height dimensions, where the channel symbol sequence at each patch is assigned with the same length value (number of symbols). This grouping strategy is inspired by the findings that the allocated rates within the kind of patch are rather similar, and thus it will not lose much flexibility even if we force the vectors within a patch to have the same length. With the spatial grouping, only one length scalar is required for multiple sequences in $\mathbf{r}_l$, and thus the overhead can be mitigated.
	
	Furthermore, as the JSCC encoder is required to encode the images into symbols of different numbers, we devise a rate attention module by incorporating the prior information in the encoding process.  The rate attention involves two inputs, the latent representation $\boldsymbol{\mu}_l$ of shape $(H,W,C)$ and the prior information $p_{\mathbf{z}_l \mid \mathbf{z}_{<l}}$. We calculate the length value for different vectors in $\boldsymbol{\mu}_l$, obtaining in total $HW$ real length values. Then, we calculate the merged length values after the spatial grouping. As we allocate the same length value to the vectors within a patch, the merged value is calculated by averaging the real length values in a sample patch. The rate attention operation accepts two inputs. The first is the concatenated information from the representation and prior information, while the second is the concatenated matrix of the real length value vector and the merged length value vector. 
	Through this operation, the index information can be fused to the encoding process, enabling the JSCC encoder block to adaptively adjust the encoding process.

	\section{ Experiments}
	
	\paragraph{Metrics and Test Datasets}
	For performance evaluation, we consider the pixel-wise peak signal-to-noise ratio (PSNR)\footnote{Codes are given in \textit{https://github.com/zhang-guangyi/HJSCC}.}. 
	We quantify the performance by considering the following datasets of different resolutions with necessary preprocessing. \textbf{Kodak}  \cite{Kodak_URL}: The dataset consists of $24$ images of resolution $512\times 768$ or $768 \times 512$. 
	\textbf{CLIC2022 Test} \cite{CLIC2022}:  The test set contains $30$ images up to size $1365 \times 2048$. 
	
	\paragraph{Benchmarks}
	To verify the performance of our image transmission models, we compare them with a range of benchmarks. First, we consider emerging deep learning-based schemes, including DeepJSCC \cite{Eirina_TCCN2019}, SwinJSCC \cite{Yang_Arxiv2023}, and the nonlinear transform source-channel coding (NTSCC) \cite{Jincheng_2023JSAC}. Second, we compare our methods with separation-based schemes widely used in real transmission systems. These include powerful image codecs such as BPG, JPEG, and JPEG2000, combined with a practical LDPC code \cite{Shahid_TCOM2013}, labeled as ``BPG + LDPC'', ``JPEG + LDPC'', and ``JPEG2000 + LDPC'', respectively. In addition to these hand-crafted image codecs, we also consider the learning-based image codecs combined with LDPC, \cite{balle_arxiv2018} and \cite{NEURIPS2018}. For the feedback JSCC schemes, we consider the most advanced JSCCformer-f \cite{HaotianWu} and the classical DeepJSCC-f \cite{deepjsccf}. In our experiments, we test various schemes across different CBRs and SNRs under AWGN channels. We take the architecture in \cite{Duan_WACV2023} as the backbone model.

	\renewcommand{\arraystretch}{1.2}

	\subsection{Comparisons and Results Analysis}
	In Fig. \ref{PSNR_CBR} (a) and (b), we evaluate the transmission performance at different CBRs under AWGN channels, with the test SNR set to $10$ dB. To ensure a reliable transmission link, we adopt $16$-order quadrature amplitude modulation (16QAM) combined with an LDPC rate of $2/3$, in accordance with the 3GPP standard. 
	%Given a bits per pixel (bpp) value, the CBR $\rho$ can be calculated as $\rho = \frac{K}{C\times H \times W}= \frac{\text{bpp}\times H \times W}{C\times H \times W \times \log_2 16 \times \frac{2}{3}}=\frac{\text{bpp}}{8}$. 
	The results indicate that our proposed HJSCC significantly outperforms the fixed-rate schemes, DeepJSCC and SwinJSCC. Compared with NTSCC, the proposed scheme achieves comparable performance. The performance gap widens with increasing image resolution and CBR. Additionally, compared to hand-crafted schemes, the proposed method achieves substantially better PSNR performance. This demonstrates the potential of utilizing HJSCC in practical wireless communication systems.
	
	\begin{figure}
		\begin{centering}	
			\includegraphics[width=0.41\textwidth]{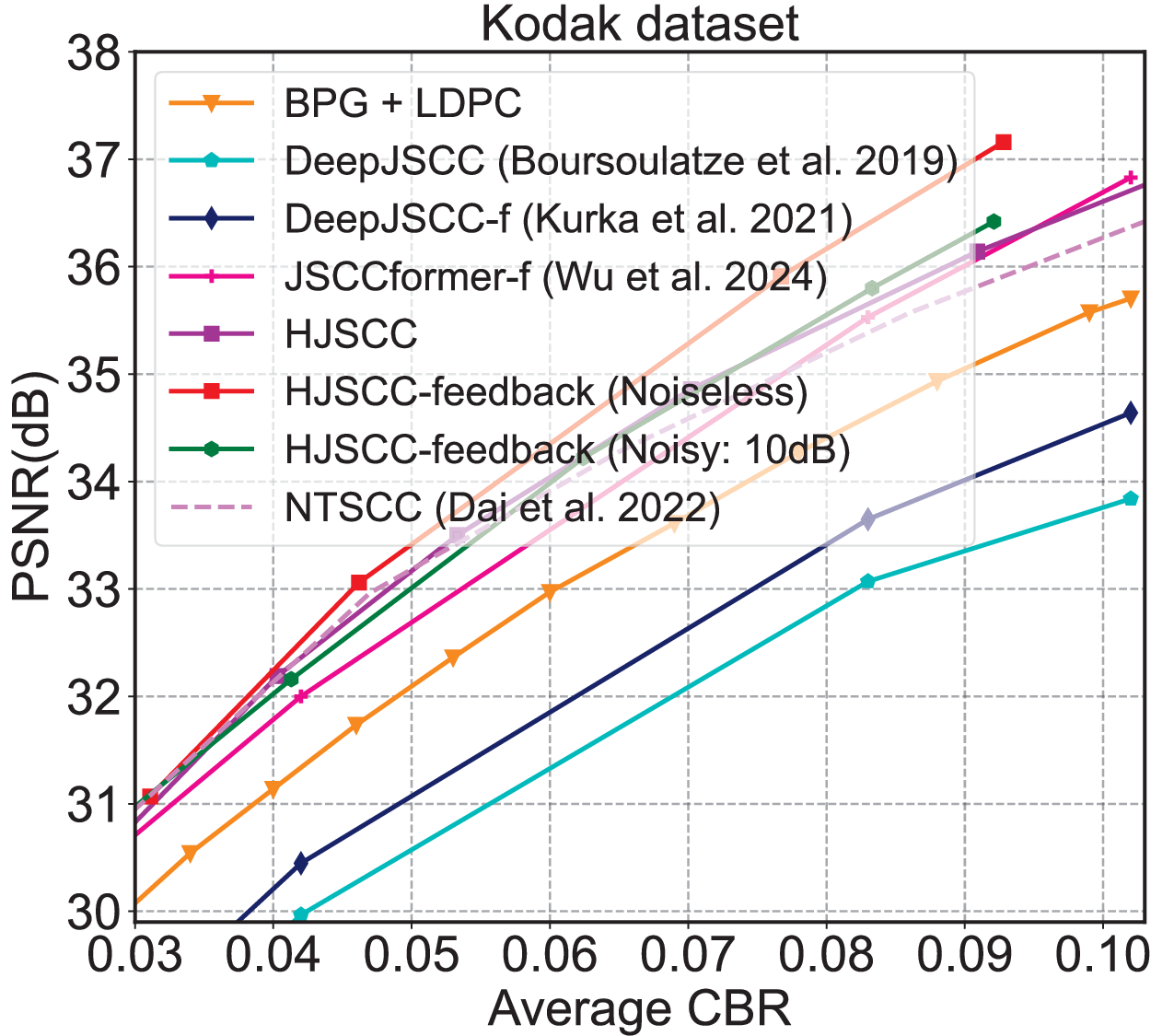}
			\par \end{centering}
		\caption{PSNR performance with feedback links.}
		\label{Feedback}
	\end{figure}

	Fig. \ref{PSNR_CBR} (c) demonstrates the transmission performance across varying channel SNR levels. To ensure fairness, the CBR for these schemes is constrained to $0.0625$. For DeepJSCC and NTSCC, the training SNR equals the testing SNR to achieve optimal performance. For the separation-based methods, we test these schemes across different channel coding rates and modulation orders to determine the optimal settings\footnote{Given a bpp value, the CBR $\rho$ can be calculated as $\rho = \frac{K}{C\times H \times W}= \frac{\text{bpp}}{C \times \log_2 M \times R_c}=\frac{\text{bpp}}{C\times \log_2 M \times R_c}$, where $M$ is the selected modulation order and $R_c$ denotes the channel coding rate. For example, $\rho=\text{bpp}/8$ when $16$QAM and rate $2/3$ are selected for SNR $=10$ dB.}, and then the image codec needs to compress the source with the resulted bits per pixel (bpp) value. Our results indicate that the proposed HJSCC significantly outperforms other schemes. Additionally, HJSCC achieves a substantial performance gain compared to the emerging method \cite{NEURIPS2018}, with the performance gap becoming more pronounced at SNR levels above $10$ dB. Furthermore, separation-based systems are known to suffer from the cliff effect, where reliable transmission cannot be maintained when the channel coding and modulation schemes fail. In contrast, the proposed HJSCC provides a graceful degradation as the SNR decreases, demonstrating its robustness in varying channel conditions.
	
	Fig. \ref{Feedback} presents the PSNR performance of different schemes, where JSCCformer-f \cite{HaotianWu} and DeepJSCC-f \cite{deepjsccf} are JSCC schemes with a feedback link. Compared with them, our proposed HJSCC shows much better PSNR performance, with a gain of about $1$ dB at high CBR region. Besides, the introduction of the feedback link also provides significantly larger gain, especially at high average CBR values. This stems from rate-adaptive capability of HJSCC-feedback as well as its generative formulation, making it the state-of-the-art JSCC schemes in the presence of a feedback link. In addition, when the feedback is noisy, we can see a slight degradation in PSNR performance. Besides, the performance improves with the feedback link quality, where a higher feedback SNR can produce better PSNR performance.

	\subsection{Ablation Studies}
	\paragraph{Ablation on the spatial grouping.} In this work, we propose a spatial grouping strategy to reduce the transmission overhead to inform the receiver of the vector length for rate matching. To show the effectiveness, we report the performance on CBR saving. Particularly, we compare the CBRs for transmitting the channel symbols $\mathbf{s}$ and the vector length information $\hat{k}$ in Table \ref{Ablation222}. All the models are optimized on ImageNet dataset. From the results, the overall CBR can be significantly reduced by the spatial merging strategy, at the cost of a slight performance degradation.

\begin{figure}
		\begin{centering}	
			\includegraphics[width=0.4\textwidth]{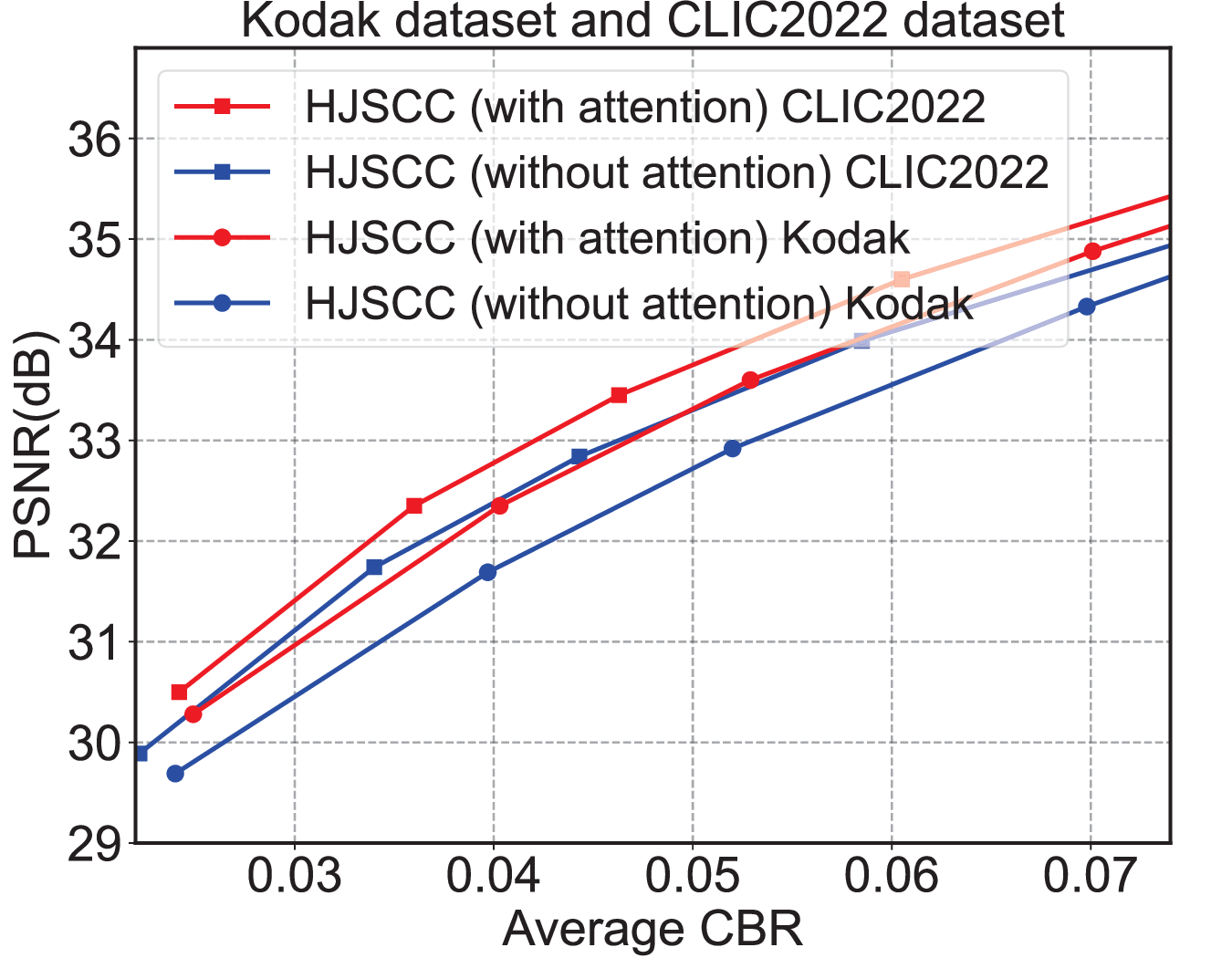}
			\par \end{centering}
		\caption{Ablation analysis on the rate attention module.}
		\label{Ablation3}
	\end{figure}

	\begin{table}[t]\footnotesize
		\centering
		\begin{tabular}{c | c | c | c | c | c}
			\toprule[1pt]
			&  $\lambda$     &   CBR     & CBR ($\mathbf{s}$)    & CBR ($\hat{k}$) & PSNR   \\     \midrule[1pt]  \midrule[1pt]
			Grouping   & $64$  & $0.0403$ & $0.0378$ & $0.0025$ & $32.03$  \\
			No grouping	 	 & $64$ & $0.0590$ & $0.0390$  &  $0.0200$  & $32.24$  \\  \midrule[1pt]
			Grouping   & $16$ &$0.0249$  & $0.0224$ & $0.0025$ &  $29.96$  \\ 
			No grouping		& $16$ &$0.0430$  & $0.0230$ & $0.0200$ & $30.13$ \\
			\bottomrule[1pt]
		\end{tabular}
		\caption{\fontsize{10pt}{17bp}Ablation analysis on the spatial grouping strategy.}  		\label{Ablation222}
	\end{table}		
	\paragraph{Ablation on rate attention module.} Moreover, we also verify the effect of the proposed rate attention module, as shown in Fig. \ref{Ablation3}. We compare the performance of using and not using this module on the Kodak and CLIC2022 datasets, over the AWGN channels, where the SNR is set to $10$ dB. From the results, we find that the proposed rate attention module significantly improves the PSNR performance over different average CBR values. This performance gain stems from the ability of the rate attention module in guiding the JSCC encoder to encode the latent representation into channel symbol vectors of different length values, demonstrating the effectiveness of this module. 	

	\section{Conclusion \& Discussion}
	In this work, we introduced a high-efficiency JSCC framework for wireless image transmission based on hierarchical VAE. Unlike conventional methods, our proposed scheme learns a hierarchical latent representation and employs multiple JSCC encoder/decoder pairs to transmit these latent representations. We formulate a novel generative formulation for HJSCC with feedback by viewing the transmission as a sampling process. By leveraging the learned prior in the JSCC encoder, our proposed HJSCC can dynamically adjust the transmission rate according to the data distribution, making it a rate-adaptive scheme compared to existing solutions.

	\section{Acknowledgments}
	The corresponding author is Yunlong Cai. This work was supported in part by the National Natural Science Foundation of China under Grant U22A2004, and in part by Zhejiang Provincial Key Laboratory of Information Processing, Communication and Networking (IPCAN), Hangzhou 310027, China.
	\bibliography{aaai25}

\end{document}